# 18F-FDG-PET dissemination features in diffuse large B cell lymphoma are predictive of outcome

Authors: Anne-Ségolène Cottereau[1,2], Christophe Nioche[2], Anne-Sophie Dirand[2], Jérôme Clerc[1], Franck Morschhauser[3], Olivier Casasnovas[4], Michel Meignan[5], Irène Buvat[2],

1. Department of nuclear medicine, Cochin Hospital, Assistance Publique Hôpitaux de Paris, Paris Descartes University, Paris, France.

2. Imagerie Moléculaire In Vivo, CEA, Inserm, Univ Paris Sud, CNRS, Université Paris Saclay, Orsay, France.

3. Univ. Lille, CHU Lille, EA 7365 - GRITA - Groupe de Recherche sur les formes Injectables et les Technologies Associées, F-59000 Lille, France.

4. Hematology department and INSERM 1231, University Hospital, Dijon, France

5. LYSA Imaging, Creteil, France.

First Author

Anne-Ségolène Cottereau

Nuclear Medicine Department, Cochin Hospital, Assistance Publique Hôpitaux de Paris, Paris Descartes University, 75014 Paris, France.

Mail: annesegolene.cottereau@aphp.fr

Tel: +33 1 58 41 41 41

ORCID iD : 0000-0002-4805-4564

Word counts for text: 4210

Figure/table count: 4/4

Reference count: 24

Short running title: PET dissemination features in DLBCL

The authors declare that they have no conflict of interest.

**ABSTRACT**

We assessed the predictive value of new radiomic features characterizing the lesion dissemination in baseline 18F-FDG PET and tested whether combining them with baseline metabolic tumour volume (MTV) could improve prediction of progression free survival (PFS) and overall survival (OS) in diffuse large B cell lymphoma (DLBCL) patients.

Methods: From the LNH073B trial (NCT00498043), patients with an advanced stage DLCBL and 18F-FDG PET/CT images available for review were selected. MTV and several radiomic features, including the distance between the two lesions that were the furthest apart ($Dmax_{patient}$) were calculated. Receiver operator characteristics analysis were used to determine the optimal cut-off for quantitative variables and Kaplan–Meier survival analyses were performed.

Results: With a median age of 46 years, 95 patients were enrolled, half of them treated with R-CHOP14 (rituximab, cyclophosphamide, doxorubicin, vincristine and prednisone), the others with R-ACVBP (rituximab, doxorubicin, cyclophosphamide, vindesine, bleomycin, prednisone), with no significant impact on outcome. Median MTV and $Dmax_{patient}$ were 375 cm$^3$ and 45 cm respectively. The median follow-up was 44 months. High MTV and $Dmax_{patient}$ were adverse factors for PFS (p=0.027 and p=0.0003 respectively) and for OS (p=0.0007, p=0.0095). In multivariate analysis, only $Dmax_{patient}$ was significantly associated with PFS (p=0.0014) whereas both factors remained significant for OS (p=0.037 and p=0.0029 respectively). Combining MTV (>384 cm$^3$) and $Dmax_{patient}$ (>58 cm) yielded 3 risk groups for PFS (p=0.0003) and OS (p=0.0011): high with 2 adverse factors (4y-PFS and OS of 50% and 53%, n=18), low risk with no adverse factor (94% and 97%, n=36), and an intermediate category with one adverse factor (73% and 88%, n=41).

Conclusion: Combining MTV with a parameter reflecting the tumour burden dissemination further improves DLBCL patient risk stratification at staging.

# INTRODUCTION

Diffuse large B-cell lymphoma (DLBCL) represents the most frequent types of lymphoid cancer, accounting for approximately 25% of non-Hodgkin lymphoma (NHL) (*1*). The current first line treatment R-CHOP – rituximab (R), a CD-20-directed monoclonal antibody, given in combination with CHOP, the standard chemotherapeutic regimen of cyclophosphamide, doxorubicin, vincristine and prednisone – is effective in 60% to 70% of patients (*2*). For the 30% to 40% of the patients who will exhibit refractory disease or relapse after initial response, the prognosis is poor. The life expectancy for patients with refractory disease or early relapse is dramatically reduced since salvage regimens lead to very modest response rates (*3,4*). A personalized approach of first line treatment might improve DLBCL patients' outcome. Interim PET (iPET) performed after 2 or 4 cycles of chemotherapy has been proposed as a tool for tailoring therapy but no therapeutic approach has proven successful to improve the prognosis of iPET positive patients. An earlier risk stratification is therefore still needed. High risk patients are not accurately identified by the current prognostic scoring systems, as the International Prognostic Index (IPI) (*5*), Revised IPI (*6*) or NCCN IPI (*7*). Over the last five years, the prognostic role of quantitative PET parameters, in particular the metabolic tumour volume (MTV), has been demonstrated in many lymphomas subtypes (*8,9*), including DLBCL (*10-12*). MTV reflects the total volume of 18F-FDG-avid tumour regions within the whole body, hence provides a more comprehensive tumour burden evaluation than previous surrogates such as lactate dehydrogenase levels. Patients with high tumour burden are at higher risk for treatment failure and shorter survival than those with low tumour burden. However, this parameter does not account for the spatial distribution of the lesions throughout the body. Yet chemokine receptor 4 (CXCR4) expression has been shown to be a marker of bad prognosis in DLBCL (*13, 14*). As CXCR4 expression mediates dissemination of DLBCL cells, our assumption was that the prognostic value of MTV might be improved by combining the tumour burden estimate with a quantitative feature reflecting the spread of the disease. The aim of this study was to define and analyse new 18F-FDG PET metrics describing the tumour dissemination, and to determine their added predictive value to MTV for DLBCL patients included in the LNH073B trial (*15*).

## MATERIALS AND METHODS

### Patients

Details and results of the LNH073B study design (the study was registered at ClinicalTrials.gov: NCT00498043) have been published elsewhere (*15*). In brief, DLBCL patients with an age adjusted IPI (aaIPI) score of 2 or 3 were randomly assigned to an induction immunochemotherapy with 4 cycles of either R-CHOP14 or R-ACVBP (rituximab, doxorubicin, cyclophosphamide, vindesine, bleomycin, prednisone). Consolidation treatment was driven by centrally reviewed PET assessment according to visual criteria after 2 and 4 treatment cycles. A baseline PET scan was mandatory, with at least one evaluable hypermetabolic lesion. Ethics approval was obtained for this trial, and all patients provided written informed consent to participate.

For the current analysis, only Ann Arbor stages 3 and 4 patients whose MTV could be computed from a baseline PET/CT scan and with at least two detectable lesions allowing distance measurement were included.

Baseline patient and disease characteristics, including individual components of the aaIPI score, progression-free survival (PFS) and overall survival (OS) defined according to the revised National Cancer Institute criteria were obtained (*16*).

### PET/CT Scanning and Quantitative Analysis

Baseline PET image data in anonymized Digital Imaging and Communications in Medicine (DICOM) format was collected for functional parameter measurements. Quality control rejected scans with burning errors in DICOM retrieval or with a delay of > 90 minutes between 18F-FDG injection and scanning.

Analysis of PET data was performed by a nuclear medicine physician (ASC) blinded to patient outcome, using the LIFEx software (*17*). MTV was calculated based on a supervised segmentation of tumour regions involving 41% SUVmax thresholding of automatically detected hypermetabolic regions. MTV was defined as the sum of every individual lesion metabolic volume.

For each lesion, the tumour lesion glycolysis was calculated as the product of the lesion volume by the SUV mean within the lesion, and the total lesion glycolysis (TLG) was obtained by summing the tumour lesion glycolysis over all lesions. The highest SUVmax of the patient over all lesions and the number of lesions were also reported. Last, several features reflecting the spatial distributions of malignant foci throughout the whole body were computed, based on distance measurements between lesions. Each lesion location was defined as the position of its center and the distances between two lesions were calculated using the Euclidian distance between their centers.

Four dissemination features were calculated in LIFEx: the distance between the two lesions that were the furthest apart ($Dmax_{patient}$), the distance between the largest lesion and the lesion furthest away from that bulk ($Dmax_{bulk}$), the sum of the distances of the bulky lesion from all other lesions ($SPREAD_{bulk}$) and the largest value, over all lesions, of the sum of the distances from a lesion to all the others ($SPREAD_{patient}$).

MTV was also calculated with FIJI software by an independent nuclear physician (MM), based on the same 41% SUVmax threshold method. Reproducibility of MTV measurements between LIFEx and FIJI software was assessed.

**Statistical Analysis**

For each PET-derived feature, Receiver Operating Characteristic (ROC) analysis was used to define the optimal cut-off for predicting the occurrence of an event (PFS or OS) by maximizing the Youden index (sensitivity + specificity-1). Sensitivity and specificity were calculated for that cut-off value. Only features with an area under the ROC curve (AUC) greater than 0.6 on PFS were retained for subsequent analyses. Survival functions were calculated by using Kaplan-Meier analyses and the survival distributions were compared using the log-rank test. Multivariate analyses involving MTV and dissemination features were performed using Cox proportional hazard models. Based on these results, a prognostic model combining MTV and a dissemination feature was built on which Kaplan Meier survival analysis was performed. Correlations between dissemination features and MTV were assessed using chi-squared tests. Mann-Whitney tests were used to test whether the patient size and MTV were significantly different in patients with low and high dissemination features. Reproducibility of MTV measurement between two operators and two different software (ASC and MM) was assessed by the Lin concordance correlation coefficient, and the interobserver agreement was assessed by using the kappa statistics.

Statistical significance was set to p<0.05. All statistical analyses were performed using MedCalc software (MedCalc Software, Ostend, Belgium).

**RESULTS**

In total, 95 patients, were included, whom clinical characteristics are summarised in Table 1.

With a median follow-up of 44 months (range 27-63 months), the 4-year PFS and OS rates for the whole group were 77 % and 85 % respectively. Twenty-two patients had a PFS event with a median of 7 months, 12 in R-CHOP group and 10 in R-ACVBP group. Thirteen patients died with a median of 13 months, 8 in R-CHOP group and 5 in R-ACVBP group. Using log rank tests, neither performance status (0-1versus 2-3) nor aaIPI (2 versus 3) were significantly associated with PFS (p=0.17, p=0.21) or OS (p=0.41, p=0.46). No significant prognostic impact of chemotherapy regimen (R-CHOP vs R-ACVBP) was observed for both PFS (p=0.69) and OS (p=0.48).

**PET Features**

Table 2 shows the descriptive statistics for the PET features and Table 3 gives the results of ROC analyses performed on each PET parameter.

Using ROC optimal cut-off, MTV was highly predictive of outcome (PFS: p=0.027 and OS: p=0.0007) (Table 4). Patients with a high MTV had a significantly worse outcome with a 4-year PFS and OS of 67% and 73% versus 84% and 95% for patients with a lower MTV (Fig. 1).

MTV calculation with two different software was reproducible, with a Lin concordance correlation coefficient of 0.85 (0.79 to 0.89) and a kappa of 0.86, suggesting an overall good agreement.

Regarding the dissemination features, ROC AUC were always greater than 0.6 for PFS, and close to 0.6 for OS (Table 3). Table 4 shows that $Dmax_{patient}$ > 58 cm, $Dmax_{bulk}$ > 43 cm, $SPREAD_{patient}$ > 1020 cm and $SPREAD_{bulk}$ > 530 cm were negative prognostic factors for PFS (p=0.0003, p=0.0003, p=0.0011, p<0.0001 respectively) and that for OS, only $Dmax_{patient}$ and $Dmax_{bulk}$ were statistically significant (p=0.0095 and p=0.023 respectively, Fig. 2).

No significant differences in height were observed between patients with low and high $Dmax_{patient}$ (p=0.96). Similarly, no significant differences in MTV were observed between patients with low and high $Dmax_{patient}$ (median of 344 $cm^3$ and 415 $cm^3$ respectively, p=0.14).

**Combination of MTV and Dissemination Features**

In multivariate Cox regression analysis including MTV and $Dmax_{patient}$, $Dmax_{patient}$ was significantly associated with PFS (p=0.0014; HR=4.3) whereas MTV was not (p=0.056; HR=2.3). For OS, both factors were significant (p=0.037; HR=4.0 for MTV and p=0.029; HR=3.7 for $Dmax_{patient}$).

Three risk categories could therefore be significantly distinguished on the basis of the presence or absence of high MTV (> 394 $cm^3$) or $Dmax_{patient}$ (> 58 cm) (p=0.0003 for PFS and p=0.0011 for OS) (Fig. 3): group 1 with no risk factor (n=36), group 2 with one risk factor only (n=41), group 3 with both (n=18), with 4-year PFS rates of 94%, 73%, and 50%, respectively and 4-year OS rates of 97%, 88%, and 53%, respectively. Group 2 vs group 3 had significantly different PFS (p=0.041) and OS (p=0.019). Group 1 vs group 2 had significantly different PFS (p=0.013) whereas OS did not reach significance (p=0.13). Figure 4 shows examples of 18F-FDG PET images (Maximum Intensity Projections) of patients belonging to groups 2 and 3.

**DISCUSSION**

Lymphoma is a group of blood cancers that develop from lymphocytes. Although most cells in the body can migrate at one or more distinct steps during their development and differentiation, the trafficking propensity of lymphocytes is unrivaled among somatic cells. In case of malignant transformation, this property allows for rapid tumor dissemination irrespective of the conventional anatomic boundaries limiting early spread in most types of cancer. Thus, the disease can spread rapidly to different parts of the body, involving lymph nodes, possibly associated with extra nodal sites (*18*).

18F-FDG-PET/CT is the current state-of-the-art imaging scan in lymphoma. Recent advances in PET imaging revealed that MTV, as a surrogate for tumour cell number, has a strong prognostic value in DLBCL, much higher than the presence of a bulk (*10,11*). Recently, this was confirmed in a large phase 3 study, GOYA, including more than 1100 patients (NCT01287741 (*19*)): MTV quartiles stratified the population in quartiles 1, 2, 3 and 4 with a three-year PFS of 86%, 84%, 78% and 66% respectively (*20*). In the present study, we demonstrated that MTV maintained its prognostic power in a cohort of advanced stage patients. Patients stage 3 or 4 were significantly stratified in two different risk categories according to their MTV. Moreover, using ROC analysis, MTV was the only significant feature on both PFS and OS. It was superior to standard features such as aaIPI for both PFS and OS. A high MTV identify 64% of the PFS events (14/22).

In this study, we introduced new radiomic features extracted from PET scans to quantify tumour dissemination. Several of these features based on distance measurement between lymphoma lesions were significant for PFS and OS in our group of stage 3 and stage 4 patients, suggesting that an advanced characterization of the lesion dissemination is relevant even among patients with an advanced disease. In particular, the distance between the two lesions that were the furthest apart, $Dmax_{patient}$, had strong predictive power for PFS and OS. A high $Dmax_{patient}$ was associated with an adverse outcome, with a 4-year PFS and OS of 55% and 69% respectively. Similarly, $SPREAD_{patient}$ and $SPREAD_{bulk}$ combining spatial spread information and the number of lesions were very significantly associated with PFS (Table 4).

$Dmax_{patient}$ is a very simple 3D feature to calculate with an intuitive interpretation. The height did not influence $Dmax_{patient}$, as height did not significantly differ between high or low $Dmax_{patient}$ groups. Given that the distance between two lesions is calculated based on their respective centers, it is not highly dependent on the lesion contours and on the fact that the contours are rather loose or tight depending on the delineation tool settings that are used. This is an asset to ensure good reproducibility.

Combining MTV and $Dmax_{patient}$ made it possible to identify a group with a poor prognosis so that clinicians might consider changing treatment. Indeed, patients with high baseline MTV (>394 cm$^3$) and high $Dmax_{patient}$ (>58 cm) had a much worse prognosis than the other patients with 4-year PFS 50% and 4-year OS 53%. This group represented 19% of the cohort and included 41% of the PFS total number of events (9/22) and 54% of the OS total number of events, making this model useful for identifying patients with poor prognosis.

In the LNH073B trial, consolidation treatment was driven by centrally reviewed PET assessment after 2 (denoted PET 2) and 4 (denoted PET 4) cycles: patients who were classified as PET 2 and PET 4 negative received standard immunochemotherapy consolidation; patients classified as PET 2 positive and PET 4 negative received 2 cycles of high-dose methotrexate (3 g/m2) and then a high-dose therapy (carmustine, etoposide, cytarabine, melphal [BEAM] or zevalin, carmustine, etoposide, cytarabine, melphalan[Z-BEAM]), followed by Autologous Stem Cell Transplantation (ASCT); PET 4 positive patients had salvage regimen followed by ASCT in responders to salvage. Despite this 18F-FDG-PET–driven consolidation strategy that might actually decrease the prognostic impact of baseline PET features, MTV and dissemination features remained significantly predictive of PFS and OS. Further studies are needed to more comprehensively establish the role dissemination features might play in lymphomas when measured at baseline and during patient monitoring. Indeed, it has been shown that dysregulated CXCR4 expression predicts disease progression in DLBCL and that CXCR4 overexpression impairs Rituximab response and prognosis of R-CHOP-treated DLBCL patients (*21*). In addition, MTV influences the Rituximab pharmacokinetics (*22*). Patients with high MTV had a lower AUC of Rituximab concentration and low AUC are associated with lower response rate, shorter PFS and shorter OS. The observed prognostic value of the dissemination biomarker we propose is consistent with the association between CXCR4 overexpression and Rituximab resistance. The relationship between CXCR4 expression and radiomic features reflecting the spread of the disease would be worth investigating to determine whether radiomic features can actually partly reflect CXCR4 expression. Imaging of chemokine receptor 4 (CXCR4) could also be helpful in this regard (*23*).

There are many molecular and imaging biomarkers proposed for baseline prognostic prediction in DLBCL among which the most recent ctDNA has been correlated with TMTV and PET response after treatment (*24*). The respective role of these new imaging and molecular biomarkers will have to be determined in large prospective studies for personalized therapeutic approaches.

**CONCLUSION**

18F-FDG-PET/CT can provide a predictive radiomic signature combining metrics reflecting tumor dissemination and tumor burden. In this study of advanced stage DLBCL patients, combining MTV and $Dmax_{patient}$ improved patient risk stratification at staging.

KEY POINTS

QUESTION: Could new radiomic features characterizing the lesion dissemination in baseline 18F-FDG PET improve survival prediction in DLBCL patients?

PERTINENT FINDINGS: In a cohort of 95 DLBCL patients from the LNH073B trial, new radiomic features extracted from PET scans based on distance measurement between lymphoma lesions were significant for PFS and OS prediction. Combining MTV with a parameter reflecting the tumour burden dissemination further improves DLBCL patient risk stratification at staging.

IMPLICATIONS FOR PATIENT CARE: Combining metrics reflecting tumor dissemination with metabolic tumor volume identified a group of high risk patients who might benefit from new therapeutic strategies.

**Figures**

Figure 1 Kaplan–Meier estimates of progression-free survival (PFS) and overall survival (OS) according to metabolic tumor volume (MTV).

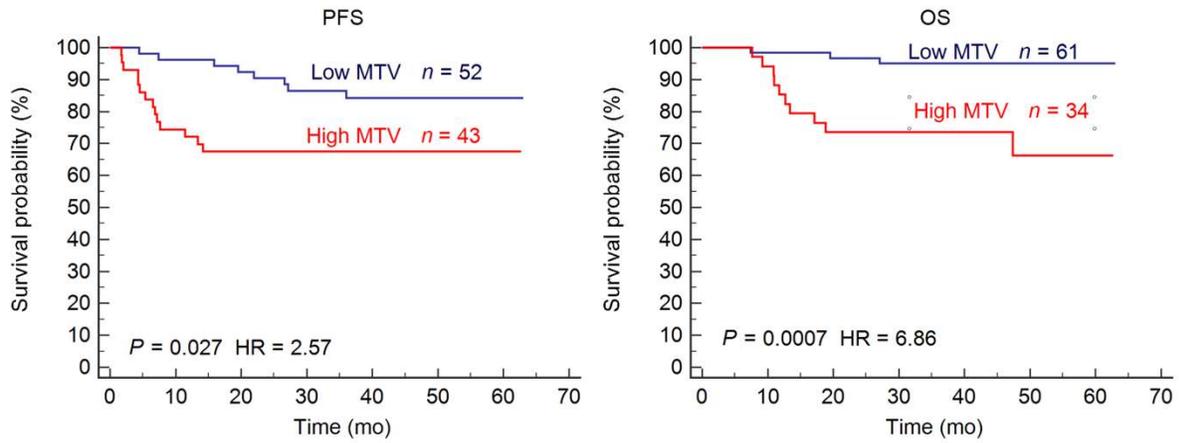

Figure 2 Kaplan–Meier estimates of progression-free survival (PFS) and overall survival (OS) according to Dmax$_{patient}$.

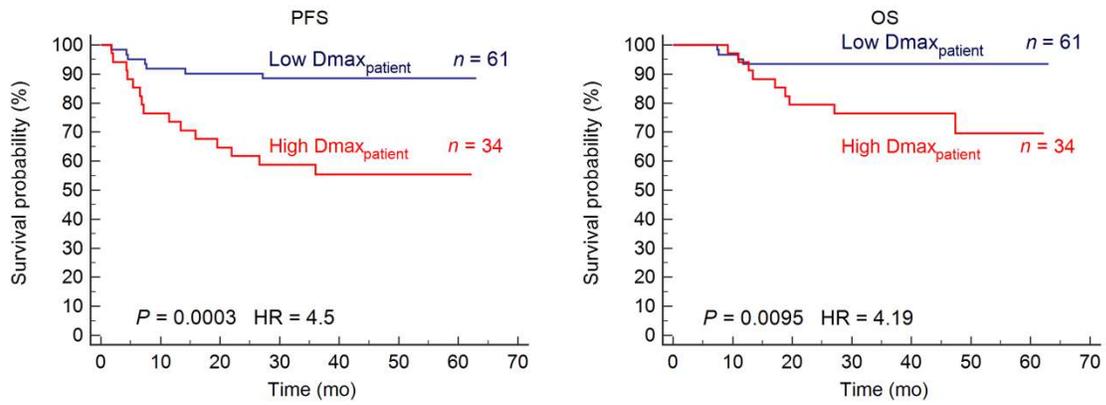

Figure 3 Kaplan–Meier estimates of progression-free survival (PFS) and overall survival (OS) according to baseline metabolic tumour volume (MTV) and $Dmax_{patient}$.

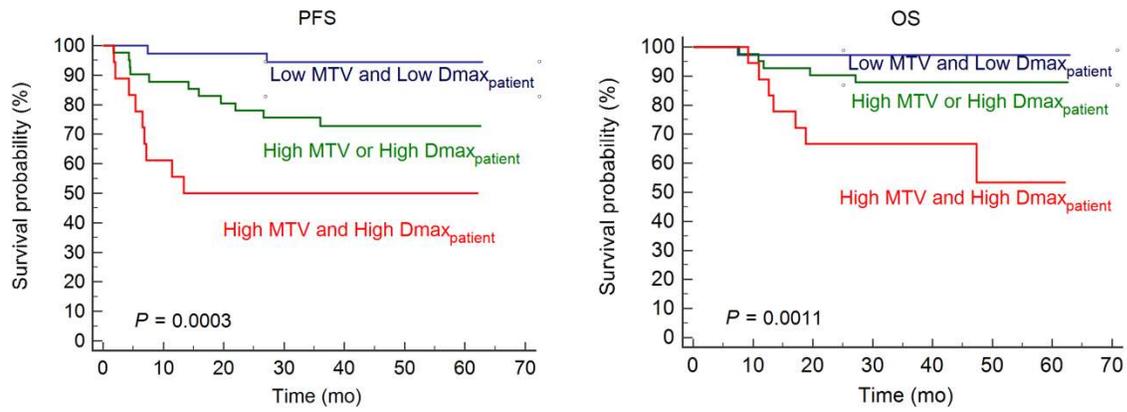

Figure 4 Examples of a patient with high MTV and low $Dmax_{patient}$ (A: group 2) and a patient with both high MTV and high $Dmax_{patient}$ (B: groupe 3).

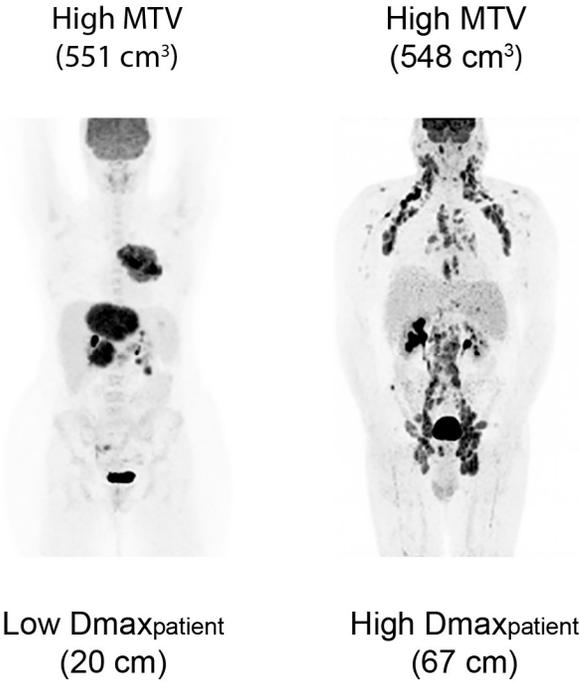

**Tables**

| Patient characteristics | n=95 (%) |
|---|---|
| **Sex** | |
| Female | 42 (44%) |
| Male | 53 (56%) |
| **Age** (median, ranges) years | 46 (18-59) |
| **Height** (median, ranges) cm | 173 (140-193) |
| ≤ 170 cm | 42 |
| >170 cm | 53 |
| **Ann Arbor stage** | |
| III | 9 (9.5%) |
| IV | 86 (90.5%) |
| **Performance status** | |
| 1 | 27 (28.4%) |
| 2 | 44 (46.3%) |
| 3 | 19 (20%) |
| 4 | 5 (5.3%) |
| **aaIPI** | |
| 1 | 3 (3%) |
| 2 | 69 (73%) |
| 3 | 23 (24%) |
| **Treatment** | |
| R-ACVBP 14 | 46 (48%) |
| R-CHOP 14 | 49 (52%) |

Table 1. Patient characteristics

| PET Parameters | *median* | *range* | *mean* | *SD* |
|---|---|---|---|---|
| **MTV** | 375 | 27-2525 | 469 | 392 |
| **SUVmax** | 20 | 4-49 | 21 | 8 |
| **TLG** | 3275 | 166-19428 | 4298 | 3323 |
| **Dmax$_{patient}$** (cm) | 45 | 7-135 | 46 | 25 |
| **Dmax$_{Bulk}$** (cm) | 32 | 7-101 | 32 | 17.5 |
| **SPREAD$_{patient}$** (cm) | 367 | 7-11915 | 798 | 1420 |
| **SPREAD$_{Bulk}$** (cm) | 205 | 7-4561 | 425.4 | 620 |
| **Nb of VOIs / patient** | 13 | 2-130 | 20 | 21 |

Table 2: Median, range, mean and standard of deviation (SD) of PET features

| PET Parameters | PFS | | | | OS | | | |
|---|---|---|---|---|---|---|---|---|
| | *AUC* | *Cut-off* | *Se* | *Sp* | *AUC* | *Cut-off* | *Se* | *Sp* |
| **MTV (cm$^3$)** | 0.64 | 394 | 68 | 60 | 0.69 | 468 | 77 | 71 |
| **SUVmax** | 0.58 | 15 | 41 | 85 | 0.53 | 23 | 46 | 71 |
| **TLG** | 0.53 | 4396 | 45 | 68 | 0.67 | 4550 | 61 | 73 |
| **Dmax$_{patient}$ (cm)** | 0.65 | 58 | 68 | 74 | 0.59 | 58 | 69 | 69 |
| **Dmax$_{Bulk}$ (cm)** | 0.63 | 43 | 54 | 82 | 0.60 | 43 | 54 | 80 |
| **SPREAD$_{patient}$ (cm)** | 0.65 | 1023 | 50 | 85 | 0.58 | 716 | 54 | 71 |
| **SPREAD$_{Bulk}$ (cm)** | 0.65 | 530 | 54 | 86 | 0.59 | 407 | 61 | 71 |
| **Nb of VOIs / patient** | 0.64 | 23 | 54 | 77 | 0.57 | 20 | 54 | 67 |

Table 3: ROC analysis of PET features, Area under the ROC curve (AUC), Sensitivity (Se), Specificity (Sp).

|  | PFS | | | OS | | |
| --- | --- | --- | --- | --- | --- | --- |
|  | HR (CI 95%) | 4y-PFS (CI 95%) | p | HR (CI 95%) | 4y-OS (CI 95%) | P |
| Low MTV | 1 (ref) | 84% (79-89) | **0.027** | 1 (ref) | 95% (92-98) | **0.0007** |
| High MTV | 2.6 (1.1-6.0) | 67% (60-74) |  | 6.9 (2.1-21.9) | 66% (56-76) |  |
| Low $Dmax_{patient}$ | 1 (ref) | 88% (84-92) | **0.0003** | 1 (ref) | 93% (90-96) | **0.0095** |
| High $Dmax_{patient}$ | 4.6 (1.9-11.2) | 55% (47-63) |  | 4.2 (1.3-13.1) | 69% (60-78) |  |
| Low $Dmax_{bulk}$ | 1 (ref) | 86% (82-90) | **0.0003** | 1 (ref) | 91% (88-94) | **0.023** |
| High $Dmax_{bulk}$ | 4.1 (1.5-11.3) | 52% (42-62) |  | 3.3 (1-11.3) | 68% (57-79) |  |
| Low $SPREAD_{patient}$ | 1 (ref) | 85% (81-89) | **0.0011** | 1 (ref) | 86% (81-91) | 0.24 |
| High $SPREAD_{patient}$ | 3.7 (1.3-10,1) | 52% (42-62) |  | 1.9 (0.5-6.8) | 78% (70-85) |  |
| Low $SPREAD_{bulk}$ | 1 | 86% (82-90) | **<0.0001** | 1 (ref) | 90% (87-93) | 0.056 |
| High $SPREAD_{bulk}$ | 4.9 (1.7-13.9) | 45% (35-55) |  | 2.8 (0.8-9.9) | 69% (59-79) |  |
| Low nb of ROIs | 1 (ref) | 85% (81-89) | **0.0052** | 1 (ref) | 87% (82-92) | 0.21 |
| High nb of ROIs | 3.1 (1.2-7.9) | 58% (49-67) |  | 1.9 (0.6-6.4) | 79% (72-86) |  |

Table 4: PET parameters associated with PFS and OS in Log-rank Cox tests